\documentclass{aa}
\usepackage{psfig, graphics}
\usepackage{txfonts}
\usepackage[dvips]{graphicx}

\title{Large-scale horizontal flows in the solar photosphere}
\subtitle{III. Effects on filament destabilization}

\author{Th.~Roudier\inst{1}, M.~\v{S}vanda\inst{2,3}, N.~Meunier\inst{1,4}, S.~Keil\inst{5}, 
M.~Rieutord\inst{1}, J.M.~Malherbe\inst{6}, S.~Rondi\inst{1}, G.~Molodij\inst{6}, V.~Bommier \inst{7}, 
B.~Schmieder\inst{6}. }
\date{Received \today  / Submitted }

\offprints{Th. Roudier}
 
\institute{
Laboratoire d'Astrophysique de l'Observatoire Midi-Pyr\'en\'ees, Universit\'e Paul Sabatier
Toulouse III, CNRS, 57 Avenue d'Azeirex, 65000 Tarbes, France
\and Astronomical Institute, Academy of Sciences of the Czech Republic, Fri\v{c}ova 298, 25165 Ond\v{r}ejov, Czech Republic
\and Astronomical Institute, Charles University, V Hole\v{s}ovi\v{c}k\'ach 2, 18200 Prague, Czech Republic
\and Laboratoire d'Astrophysique, Observatoire de Grenoble, Universit\'e Joseph Fourier, 
BP 53, 38041 Grenoble cedex 9, France
\and National Solar Observatory, Sacramento Peak, Sunspot, NM 88349, USA
\and LESIA, Observatoire de Paris, Section de Meudon, 92195 Meudon, France
\and LERMA, Observatoire de Paris, Section de Meudon, 92195 Meudon, France
}

\begin{document}

\authorrunning{Roudier et al.}
\titlerunning{Large-scale horizontal flows effects on filament destabilization}

 \abstract{}
{ We study the influence of large-scale photospheric motions on the destabilization of an eruptive filament, 
observed on October 6, 7, and 8, 2004, as part of an international observing campaign (JOP~178).}
{Large-scale horizontal flows were invetigated from a series of MDI full-disc Dopplergrams and magnetograms. 
 From the Dopplergrams, we tracked supergranular flow patterns using the local correlation tracking (LCT) technique.
We used both LCT and manual tracking of isolated magnetic elements to obtain horizontal velocities from magnetograms.}
{ We find that the measured flow fields obtained by the different methods are well-correlated on large scales. 
The topology of the flow field changed significantly during the filament eruptive phase,
suggesting a possible coupling between the surface flow field and the coronal magnetic field. We measured an
increase in the shear below the point where the eruption  starts and a decrease in shear after the eruption. 
We find a pattern in the large-scale horizontal flows at the solar surface that interact with differential rotation.}
{ We conclude that there is probably a link between changes in surface flow and the disappearance of the eruptive filament.}

\keywords{The Sun: Atmosphere -- The Sun: Filaments -- The Sun: Magnetic fields}
\maketitle

\section{Introduction}

  Dynamic processes on the Sun are linked to the evolution of the magnetic field as it is influenced by
the different layers from the convection zone to the solar atmosphere. In the photosphere,
magnetic fields are subject to diffusion due to supergranular flows and to the large-scale motions of
differential rotation and meridional circulation. The action of these surface motions on magnetic fields 
plays an important role in the formation of large-scale filaments (Mackay and Gaizauskas 2003). In particular, 
the magnetic fields that are transported across the solar surface can be sheared by dynamic surface
motions, which in turn result in shearing of the coronal field. This corresponds to the formation of coronal
flux ropes in models, which can be compared with H$\alpha$ filament observations (Mackay and van Ballegooijen 2006b).
 Many theoretical models try to reproduce the basic structure and the stability of
filaments by taking  surface motions into account, as quoted above. These models predict that
magnetic flux ropes involved in solar filament formation may be stable for many days and then
suddenly become unstable, resulting in filament eruption. Observations show that twisting motions are
a very common characteristic of eruptive prominences (see for example Patsourakos and Vial 2002). However, 
it is still unknown whether the magnetic flux ropes emerge already twisted or if it is only the
photospheric motion that drives the twisting of the filament magnetic
field. The destabilization of the filament can also be linked to oscillations (Pouget 2006). 
Therefore, it remains uncertain as to the mechanisms that drive filament disappearance.
 Destabilization can come from the interior of the structure or by means of an outside flare.

  In a previous paper (Rondi et al. 2007, hence forth Paper I), local horizontal photospheric flows were measured
at high spatial resolution (0.5\arcsec) in the vicinity of and beneath a filament before and during the 
filament's eruptive phases (the international JOP178 campaign). It was shown that the disappearance of the filament 
originates in a filament gap. Both parasitic and normal magnetic polarities were continuously swept into the gap
by the diverging supergranular flow. We also observed  the interaction of opposite 
polarities in the same region, which could be a candidate for initiating the destabilization of the filament by 
causing a reorganization of the magnetic field.

 In this paper we investigate the large-scale photospheric flows at moderate spatial resolution (2\arcsec) 
beneath and in the vicinity of the same  eruptive filament. The observation and coalignment between data
from various instruments are explained in Section~2. In Section~3, we describe the different methods of 
determining the flow field on the Sun surface. The large-scale flows associated with the filament are shown Section~4.
 The properties of these flows before and after the filament eruption are described in Section~5.
In Section~6, we investigate the topology of horizontal flows in the filament area over the 3 days 
around the filament eruption. A discussion of the results and general conclusions can be found in Section~7.

\section{Observations}

 During three consecutive days of the JOP 178 campaign, Oct 6, 7, and 8, 2004
(http://gaia.bagn.obs-mip.fr/jop178/index.html), we observed the evolution of a filament that was close to 
the central meridian. We also observed the photospheric flows directly below the filament and in its immediate 
vicinity. The filament extends from $-$5\degr{} to $-$30\degr{} in latitude. A filament eruption
was observed on October 7, 2004, at 16:30~UT at  multiple wavelengths from ground and space instruments.
The eruption produced a coronal mass ejection (CME) at approximately 19:00~UT that was observed with LASCO-2/SOHO
and two ribbon flares observed with the SOHO/EIT.  MDI/SOHO longitudinal magnetic field and Doppler
velocity were recorded  with a cadence of one minute during the 3 days (see Table 1). 
The Air Force O-SPAN telescope located at the National Solar Observatory/Sacramento Peak provided a full-disc  
H$\alpha$ image every minute. The pixel sizes were 1.96\arcsec{} for
MDI magnetograms and Dopplergrams and 1.077\arcsec{} for O-SPAN  H$\alpha$ images.
Table 1 summarises the characteristics of all the observations of JOP~178 used in our analysis.

\begin{table*}
\caption{Datasets on October 6, 7, and 8, 2004.}
\label{table:1}
\begin{center}
\begin{tabular}{cccccc}
\hline \hline
Telescope & Datatype & Field of view &  Pixel size &   Cadence &   Time~ U.T.\\ \hline

ISOON & $H\alpha$ & full-disc & $1.077\arcsec$   &  1 min & 14:05--22:35  October 6\\\
 &  & & &  &  13:30--20:31  October 7 \\
 &  & & &  &  13:37--22:35 October 8 \\
\\
MDI/SOHO & magnetogram & full-disc & $1.96\arcsec$   & 1 min & 20:49--23:49   October 6\\\
  & Doppler velocity & &  &  &  ~9:44--22:50  October 7  \\
  &  & & & 1 min & ~6:56--12:52  October 8\\
  &  & & & 1 min & 16:09--19:33 October 8\\
\\
\hline
\end{tabular}
\end{center} 
\end{table*}

We coaligned all the data obtained by the different instruments (see Paper I for a complete description of the 
co-alignment procedure). Our primary goal was to derive
the horizontal flow field below and around the filament. Co-alignment between SOHO/MDI magnetograms and O-SPAN
data was accomplished by adjusting the chromospheric network visible in H$\alpha$ (O-SPAN) and the amplitude of 
longitudinal MDI magnetograms to an accuracy of one pixel (1.96\arcsec).   

\begin{figure}
\centerline{\psfig{figure=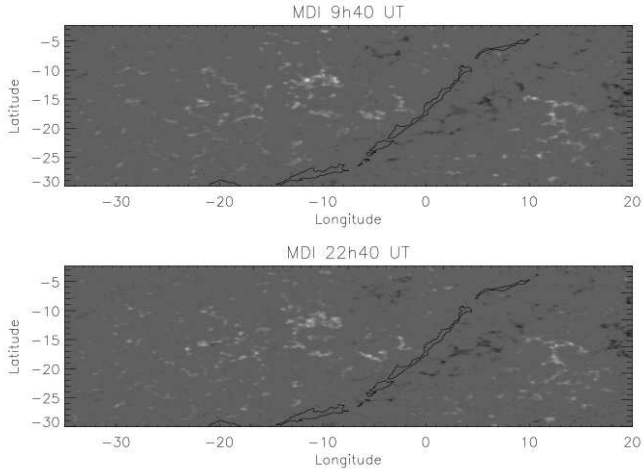,width=9 cm}}
\caption[]{ MDI longitudinal magnetic field data on October 7, 2004, at the beginning of the sequence 9:40~UT 
and at its end 22.40~UT. In order to see the magnetic field evolution, we overplot a contour of
filament observed at 13:30~UT in both figures.}
\label{context8}
\end{figure}

The general magnetic context before and after the filament eruption is shown Fig.~\ref{context8}.

\section{ Determination of the photospheric flows}

 Horizontal flows on the solar surface  may be measured through the proper motion of the plasma
or by the effects of the plasma on the magnetic structures.

 \subsection{Dopplergram processing}

 In order to map the horizontal component of the large-scale photospheric plasma velocity fields, we 
applied local correlation tracking (LCT; November 1986) to a set of full-disc Dopplergrams obtained 
by the MDI instrument onboard SoHO. The aim of this method is to track the proper motion of supergranules 
that are clearly detectable on Dopplergrams  everywhere except for the disk centre. The 
Dopplergrams were processed following the procedure described in \v{S}vanda et al. (2006) with slight modifications 
for our data. Hereafter we refer to this method as \emph{LCT-Doppler}.
  Initially we suppress the solar $p$-mode oscillations. Using a weighted temporal average (see Hathaway 1988) 
described by the formula:
\begin{equation}
w(\Delta t)=  \exp\left[{\frac{b^2}{2a^2}}\right] \left(1+\frac{b^2-(\Delta t)^2}{2a^2}\right) - \exp\left[{\frac{(\Delta t)^2}{2a^2}}\right]
\end{equation}
where $\Delta t$ is the time interval between a given frame and the central one (in minutes), $b=16$~min and $a=8$~min.
The normalized version of this function has been applied to the data.
This filter reduces the amplitudes of the solar oscillations in the 2--4 mHz frequency band by a factor of
more than five hundred. The oscillations in each frame were reduced using a window of 31 successive frames.
The different time series were tracked using the Carrington rotation rate (with an angular velocity of
13.2 degrees per day), so that all the frames have the same heliographic longitude of the central meridian
($l_0=62.24\,^\circ$). The tracked data were remapped into a sinusoidal pseudocylindrical coordinate system
(also know as Sanson-Flamsteed grid) to reduce the distortion of structures in the Dopplergrams caused by
the geometrical projection to the disc. The sinusoidal projection is suitable to describe the behaviour on the
large scales. Tracked and remapped time series then undergo a $k$--$\omega$ filtering 
with cut-off velocity of 1500~m\,s$^{-1}$ to suppress the noise coming from the groups of granules and the 
change of contrast of supergranular structures due to the solar rotation. The individual frames
were apodized by  10\% using a smooth function, the same apodization took place in the temporal domain. 
The resulting data series of tracked, remapped and filtered frames were then ready for tracking. 

The LCT method applied to full-disc Dopplergrams is characterised by a Gaussian correlation window 
($FWHM=60^{\prime\prime}$) and a time lag between correlated frames of 1~hour (basically 60 frames).
 In all cases, one half of the intervals were before the eruption and the second half after the eruption.
All the pairs of correlated frames in the studied intervals were averaged to increase the signal to
numerical noise ratio.

 \subsection{Magnetogram processing}

 The second method by which we determine  motions on the solar surface used the full-disc magnetograms
obtained by MDI/SoHO. To reduce the distortion of structures seen in the magnetograms caused by the
geometrical projection to the disc, we applied Sanson-Flamsteed grid projection. To measure the differential
motions of features on the solar surface, the data were aligned on a band along the equator. Due to the numerous 
magnetic structures to be tracked manually, the field of view was limited to  $-$35\degr{} to 20\degr{} in longitude 
and $-$3\degr{} to $-$30\degr{} in latitude. The displacement of the longitudinal magnetic structures 
visible on the magnetograms (both positive and negative) was determined with two different methods:

\begin{enumerate}
\item The first approach, hereafter named \emph{manu-B}, was to manually locate each magnetic structure 
in each magnetogram to determine its trajectory and its horizontal velocity. Once the velocities in the 
field of view were determined, as the magnetic structures do not sample the field of view unifomly, we applied a
reconstruction of the velocity field based on multi-resolution analysis described by Rieutord et al. (2007),
which allows us to limit the effects of the noise and error propagation.

\item The second approach, named \emph{LCT-B}, was to apply the LCT on magnetic structures with absolute values greater 
than 25 Gauss to reduce the noise. The LCT method applied to these magnetograms used a Gaussian correlation
window ($FWHM=60^{\prime\prime}$) and a time lag between correlated frames of 1~hour.
\end{enumerate}

 The horizontal flow field measured using the plasma (Dopplergrams) differs from that measured using
 magnetic features. This results because the magnetic field structures are not actually passive scalars; they 
can interact with the plasma and are also constrained by their interactions in the upper atmosphere. 
It is therefore not surprising to observe small differences in the velocity fields derived from 
the various tracers (Dopplergram or magnetic structures).

\section{Photospheric flow pattern below and around the filament}

\begin{figure}
\resizebox{9cm}{!}{\includegraphics{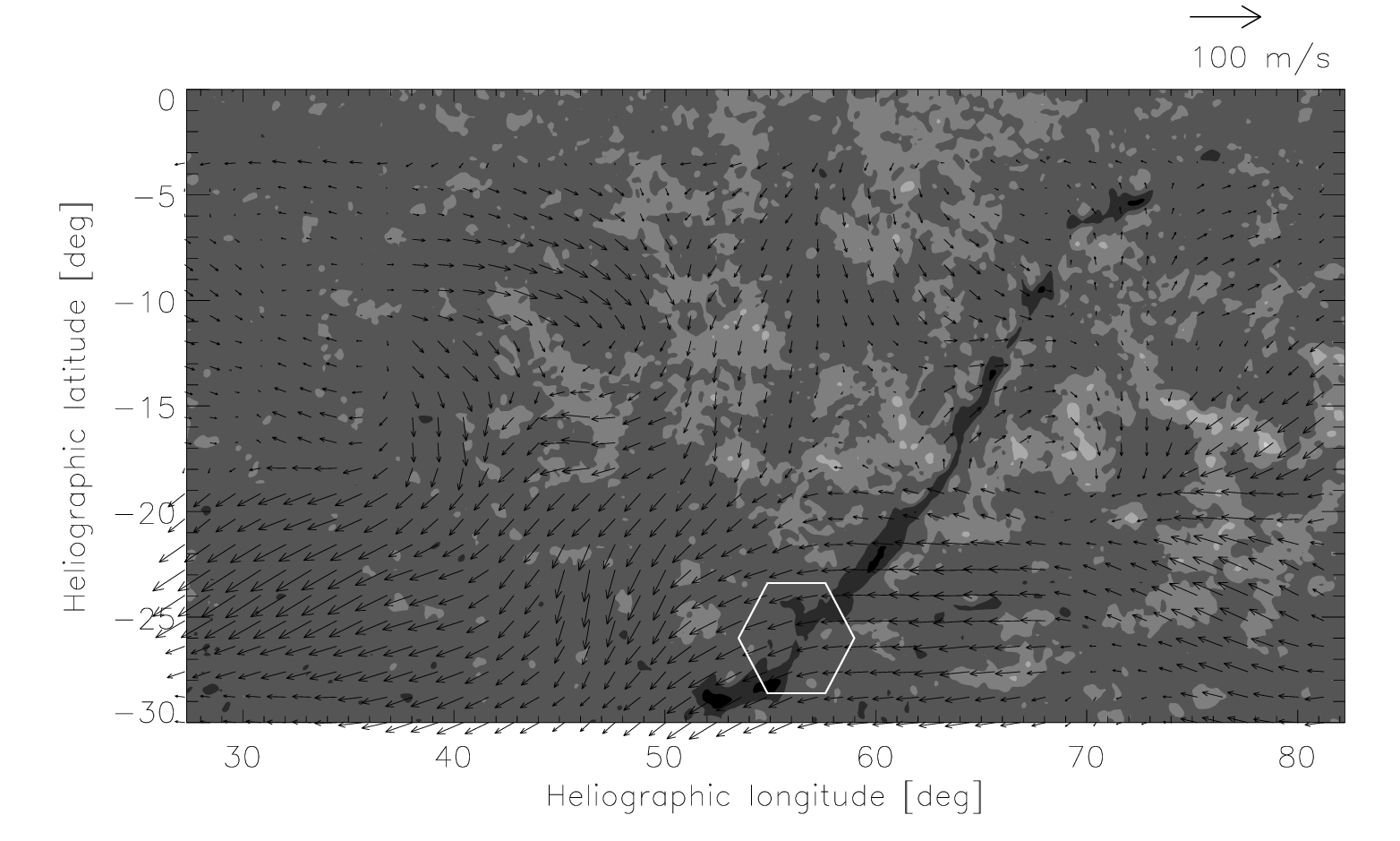}}\\
\resizebox{9cm}{!}{\includegraphics{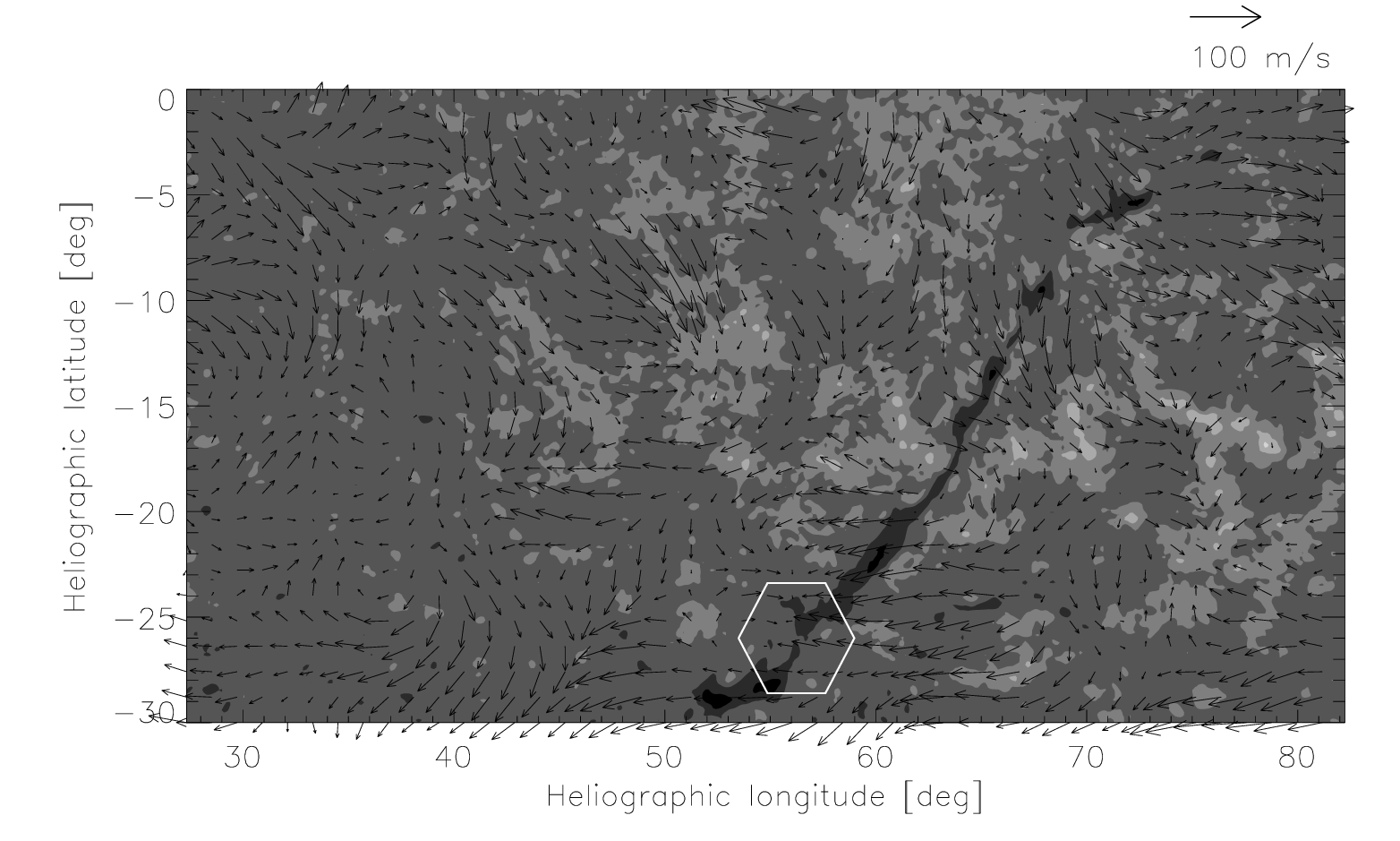}}\\
\resizebox{9cm}{!}{\includegraphics{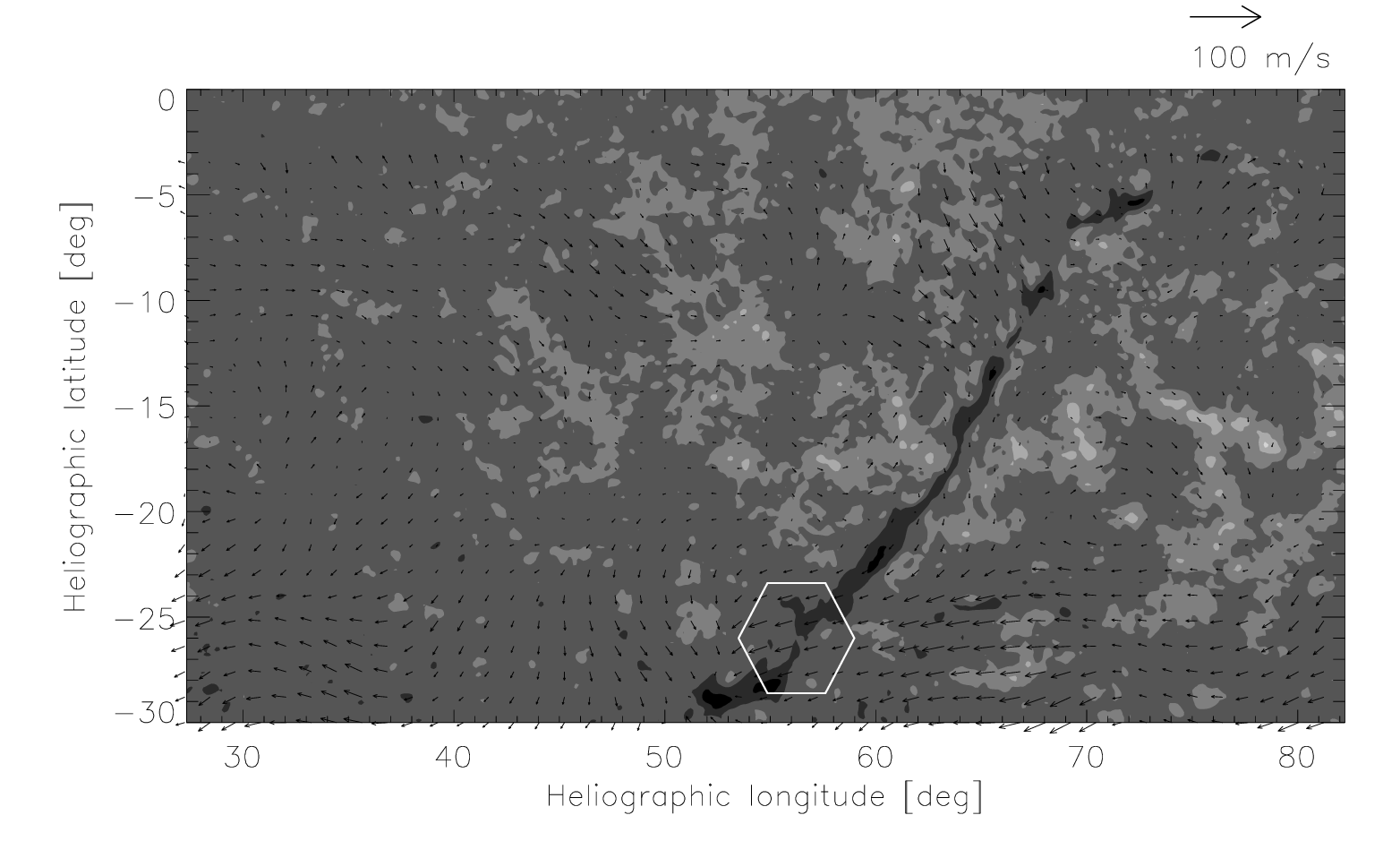}}
\caption[]{The horizontal photospheric flow fields that were derived using the various methods described in the text. 
Upper panel -- \emph{manu-B} method; 
middle panel -- \emph{LCT-Doppler} method; and bottom panel -- \emph{LCT-B} method. The filament as observed by ISOON 
in H-halpha on October 7, 2004, is superimposed. The hexagon indicates the location where the filament eruption started.}
\label{context3}
\end{figure}

In this Section we describe the flows associated with the filament eruption, with particular emphasis on the 
filament evolution and properties of the the mean East-West (zonal) velocities.

\subsection{Flow fields}

The flow fields in the vicinity of the filament obtained from 13-hour averages of velocities found using
the three different methods described above are shown in Fig.~\ref{context3}. Inspection of these 
maps shows that all three methods provide similar large-scale velocity patterns. The amplitudes of 
the flows are given in Table~2. The LCT method applied to magnetic structures with amplitude greater
than 25 Gauss appears to underestimate the amplitude of the flows obtained from the two other methods. 
This is partially because of to the large window ($FWHM=60^{\prime\prime}$) used in the LCT method, combined with 
the uneven distribution of the magnetic structures in the field of view. The correlation between zonal component, 
$v_x$, between the 
three methods is between 0.48 and 0.40, while the correlation for meridian component, $v_y$, is between 
0.21 and 0.11. The low correlation coefficients for the meridian component probably occurs because
 the main direction of the flows in our field of view is zonal and theamplitude of the meridian component is small. 
We estimate that the directional error in our measurements of the velocities is about $\pm$5\degr{}.
A small error in determining the direction of the flow, for example 10\degr{}, can affect the amplitude of the 
meridian component by a factor two. This error would greatly decrease the correlation in the meridian component
of the flow. However, as seen in Fig.~\ref{context3}, the general trend is similar in the velocity fields resulting from
the three different methods. In particular, the north--south stream that disturbs the differential rotation around  
 $-$25\degr{} is easily visible, and most of the large-scale features of the vector orientations can be identified.
We note that the lowest correlation coefficient  is found between \emph{LCT-B} and the other methods \emph{manu-B}, 
\emph{LCT-Doppler}. This results when the  \emph{LCT-B} method clearly underestimates the amplitude flows. In 
agreement with the previous results of Schuck (2006), we conclude that this method is probably not suitable
for accurately estimating horizontal velocities of magnetic footpoints.

 The correlation is higher in the longitudinal region between heliographic coordinates of $-$5\degr{} 
and 20\degr{} where it is between  0.58 and 0.41 for $v_x$  and between 0.34 and 0.17 for $v_y$.
 In this region, the large-scale flows are well-structured and show both converging and diverging velocity patterns. 
We observe in particular a large-scale stream in the north--south direction parallel to the filament located about 
10\degr{} to the east and between $-$20\degr{} and $-$30\degr{} in latitude and 58\degr{}and 47\degr{} in longitude. 
This flow stream is clearly visible in Fig.~\ref{context3} (upper and middle panels), and its dynamics can be seen
at http://gaia.bagn.obs-mip.fr/jop178/oct7/mdi/7oct-mdi.htm. Around $-$20\degr{} in latitude, the velocities
of the differential rotation amplitude appears to dominate. However, in both measurements we observe
that the north--south large-scale stream on the eastern edge of the filament disturbs the regular differential
rotation. The north--south stream is located close to  where the filament eruption begins
(longitude $l=56^\circ$, latitude $b=-26^\circ$ in Carrington coordinates). In the \emph{manu-B} and \emph{LCT-B} 
methods, the stream appears closer to the filament. The amplitudes of the southward motions are  
\emph{manu-B} 31.2~m\,s$^{-1}$, \emph{LCT-Doppler} 40~m\,s$^{-1}$, and \emph{LCT-B} 13~m\,s$^{-1}$ which are close to 
the mean observed flows (see Table 2).
 Below the latitude of $-$20\degr{}, the combination of
differential rotation and the north--south stream cause opposite polarities to move closer together, which
strongly increases the tension in the magnetic field very close to the starting point of the filament eruption. Our 
measurements show a good agreement between the \emph{manu-B} and \emph{LCT-Doppler} methods.

\begin{figure}
\centerline{\psfig{figure=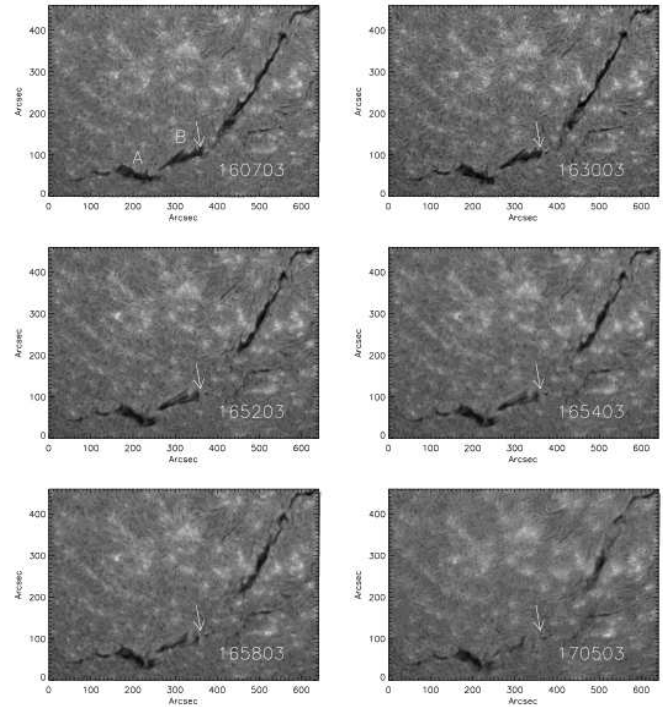,width=9 cm}}
\caption[]{ Evolution of the filament during its eruption around 16:30~UT on October 7, 2004.
The arrow indicates a fixed position for all the subframes. A and B denote two parts of the filament.}
\label{context13}
\end{figure}

 \subsection{Filament evolution}

  The filament's evolution can be seen in  Fig.~\ref{context13}. The north--south stream flow visible
in  Fig.~\ref{context6} (left) crosses over the part of the filament labeled A. The arrow on Fig.~\ref{context13}
indicates the same fixed point (325\arcsec,167\arcsec) in all of the subframes. We observe a general southward
motion of both the A and B segments of the filament. More precisely, we measure a tilt of these
two filament segments at the point of their separation. Between 16:07~UT and 16:58~UT, the long
axis of segment A  rotates by an angle of 12\degr{} (clockwise) relative
to its western end, and the long axis of segment B of the filament rotates by an angle of 5.5\degr{}
(clockwise) relative to its western end.  These rotations are compatible with the surface flow shown in
Fig.~\ref{context6} (left) and in particular the north--south stream flow. 
 We did not find a singular pivot point where differential rotation did not displace the filament with respect
to the flare location (Mouradian 2007) for the present filament. The southern segment of the filament does not reform
after the eruption. This sudden disappearance shows there was an important
change in the sun and not only in the solar atmosphere (Mouradian 2007).

\begin{table}
\caption{Velocity amplitude in m\,s$^{-1}$.}
\label{table:2}
\begin{tabular}{cccc}
\hline \hline
   &  \it manu-B & \it LCT-Doppler &  \it LCT-B  \\ \hline

mean velocity & 43.6 & 28.7 &  10.6  \\
maximum   velocity &  111.1 & 87.0 &  32.5   \\
minimum   velocity &  0.9 & 0.26 &   0.04  \\

\\
\hline
\end{tabular}
\end{table}

\begin{figure}
\centerline{\psfig{figure= 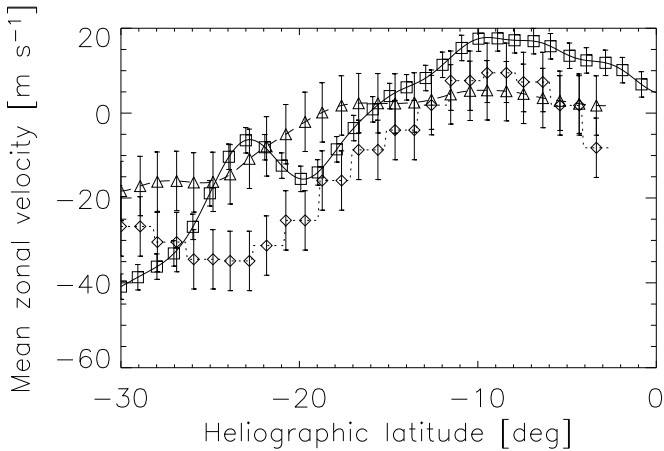,width=10cm}}
\caption[]{Mean zonal velocity is computed between $-$35\degr{} and $+$20\degr{} in longitude
and $-$30\degr{} to 0\degr{} in latitude in the Carrington system. \emph{Dash-dot line and diamonds} 
are zonal velocity for the motions measured by hand on B long  maps (\emph{manu-B} method). 
\emph{Solid line and squares} represents are zonal velocity for the LCT applied 
on Doppler images (\emph{LCT-Doppler} method).
\emph{ The dashed line and triangles} represents mean zonal velocity for the LCT method applied on B long 
maps (\emph{LCT-B} method).}
\label{context4}
\end{figure}

\subsection{Mean zonal velocities}

  Figure~\ref{context4} show plots of the mean zonal velocities resulting from the three
different velocity  determination methods as functions of latitude.  The mean zonal velocities from 
the \emph{manu-B} method clearly show the differential rotation profile with a plateau around $-$25\degr{}, 
which corresponds to the effects of the north--south stream discussed above. A similar profile, but 
with a lower amplitude, is visible in the mean zonal velocities obtained from the \emph{LCT-B} method. These two methods
measure the displacement of the magnetic structures on the Sun's surface. The \emph{LCT-Doppler} method, which measures
the motion of the photospheric plasma, shows a mean zonal velocity profile in which 
differential rotation is clearly visible, along with a strong secondary maximum visible at $-$23\degr{} of latitude.
 The secondary maximum indicates a decrease in the amplitude of the $v_x$ component because
the flows in that region are oriented more in the north--south direction. That is partly due
to the presence of the north--south stream described above and to the local organisation
of the flow. In particular, converging and diverging flows in this region seem to have a North--South 
orientation.

\begin{figure}
\centerline{\psfig{figure=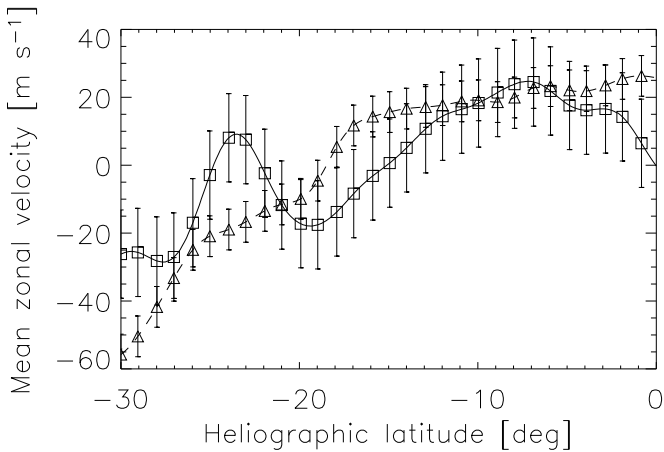 ,width=10cm}}
\caption[]{Profiles for 7 Oct in different longitudinal belts
\emph{solid line and squares} $-$25\degr{} to $-$17\degr{} degrees with respect to the centre of the
disc (not Carrington coordinates), \emph{dashed line and triangles} for 0\degr{} to 20\degr{}. }
\label{context14}
\end{figure}

 To distinguish the effects of the north--south stream from differential rotation,
we computed the  mean zonal velocities from $-$25\degr{} to $-$17\degr{} in longitude centred on the region 
where the north-south stream is present and from 0\degr{} to 20\degr{}
in longitude where, differential rotation dominates, with respect to the disc centre velocity. 
The  mean zonal velocities in the longitudinal belt, where the  north--south stream is visible, clearly exhibits 
a secondary maximum (Fig.~\ref{context14}) indicating that the solar rotation rate at this location is closer to
that of the equator. The  mean zonal velocities computed in the standard differential rotation
belt show the classic latitude profile with a constant decrease down to the low latitudes.
As a consequence, the plasma in the north--south stream, which transports magnetic structures, rotates
faster at about $-$23\degr latitude than do the
magnetic structure located in the belt of longitude between 0\degr{} to 20\degr. The combination of these different 
surface motions (stream and differential rotation) tends to bring together fields with opposite polarities, and
this in turn constrains the magnetic field lines.

 We noted that the location of the starting point of the filament eruption is around $-$26\degr{} 
in latitude, which is very close to the secondary maximum in the mean zonal velocity. Thus surface
motions that bring  opposite polarities together  may play a role in triggering the filament eruption.

\section{Flow fields before and after the eruption}

\begin{figure*}
\resizebox{9cm}{!}{\includegraphics{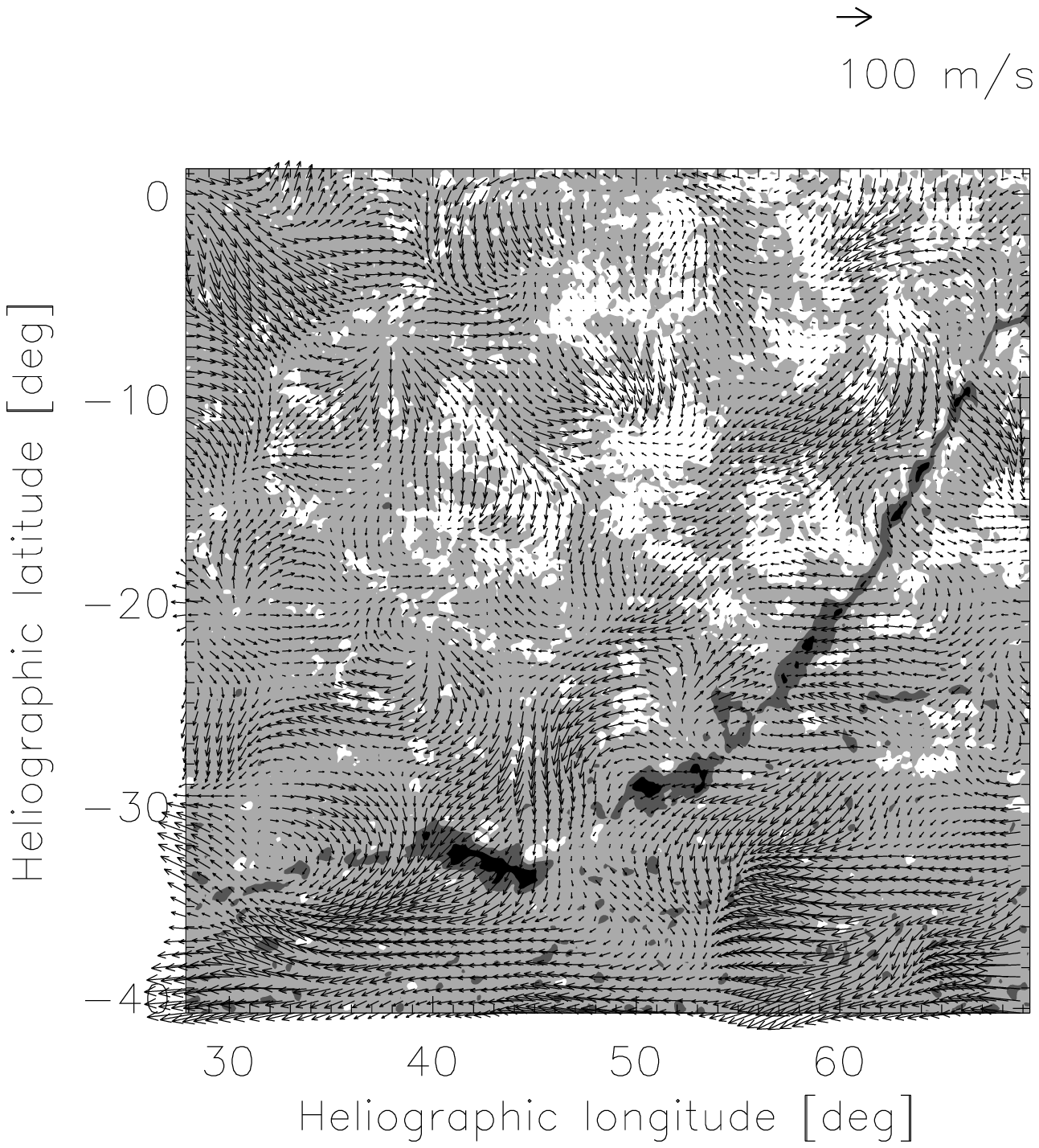}}
\resizebox{9cm}{!}{\includegraphics{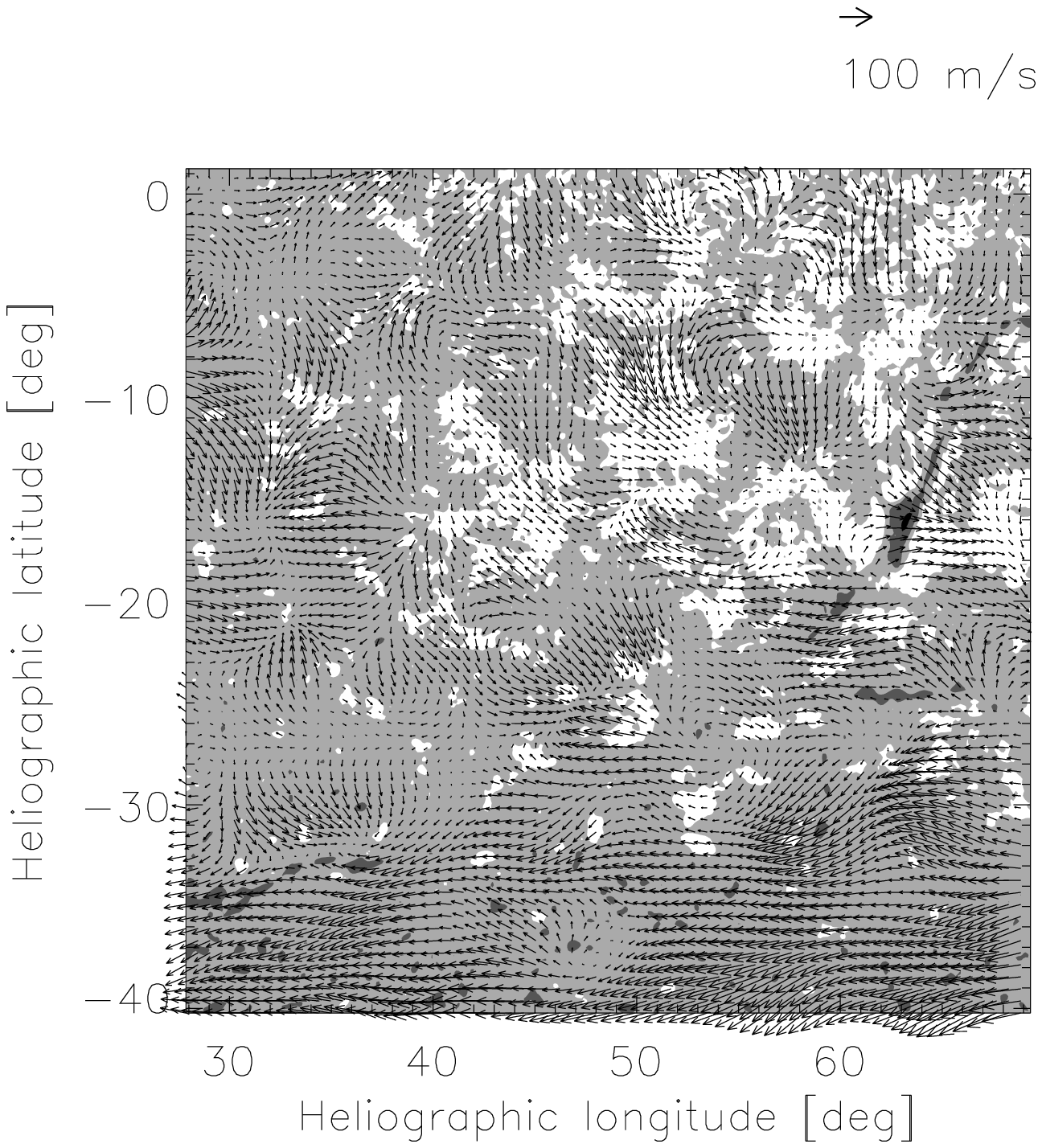}}\\
\caption[]{ Horizontal motions measured by \emph{LCT-Doppler} before (left) and 
after (right) the eruption over a wide field of view. The filament observed by ISOON on 2004 October 7 
at 13:30~UT is superimposed}
\label{context6}
\end{figure*}

\begin{figure*}
\resizebox{10.4cm}{!}{\includegraphics{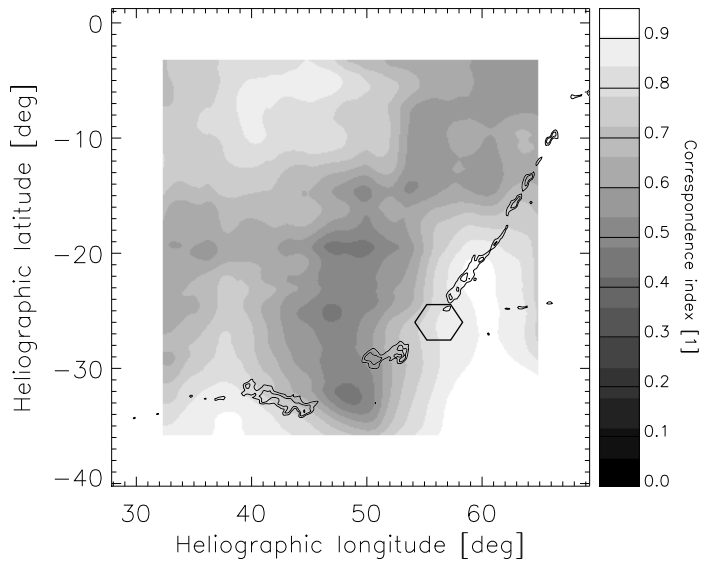}}\\
\caption[]{  Magnitude-weighted cosine map computed between flows before and after of the
eruption of the filament. White areas indicate a good agreement between the fields before and after
the filament eruption. Black areas indicate the location where a lot of changes between horizontal 
velocities take place. The filament observed by ISOON on 2004 October 7 at 13:30~UT is superimposed.
The hexagon indicates the location where the filament eruption started.}
\label{context61}
\end{figure*}

 In this section we discuss the properties of the flow field just before and after
the filament eruption, which occurred at about 16:30 UT. Due to the length of the sequence 
and because it is easier to accurately track the flow using Doppler images than it is to track the small 
number of magnetic features above 25 Gauss, only the measurements obtained with the \emph{LCT-Doppler} 
are used in this section.

At the point where the filament eruption begins ($l=56^\circ$, $b=-26^\circ$ in Carrington coordinates),
we detected a steepening of the gradient in the differential rotation curve. During the eruption, the gradient
flattens out and a dip forms. While differential rotation curves describe mean zonal velocities over most of the
disc, their change in gradient signifies a change in the stretching forces
influencing the magnetic field loops over the area under study. We can express the surface rotation 
as an even power of $ \sin\phi$ : $R=A + B \sin^2\phi+ C \sin^4\phi$, where $A$ is the angular velocity rate
of the equatorial rotation and $\phi$ the heliographic latitude. From the data we find that the constant values 
(with their errors in parentheses) are before the eruption $A=13.375(0.010)$, $B=-1.46(0.10)$, $C=-1.42(0.20)$ and 
after the eruption $A=13.404(0.010)$, $B=-1.78(0.10)$, $C=-1.24(0.20)$.
All of the rates are synodical in deg\,day$^{-1}$. The full-disc profiles did not change significantly from before to 
after the eruption. For example, for a latitude of $-$30\degr{} the zonal velocity has values of 12.92 (resp. 12.88) deg\,day$^{-1}$ 
($-$34~m\,s$^{-1}$, resp. $-$39~m\,s$^{-1}$ in the Carrington coordinate system), for a latitude of $-$20\degr{} 
the values are 13.18 ($-$2~m\,s$^{-1}$), resp. 13.18 ($-$3~m\,s$^{-1}$),~deg\,day$^{-1}$. Although the parameters of 
the smooth fitted curve did not change too much, the local residual with respect to the smooth curve changed at the 
latitude where the filament eruption starts.

Figure~\ref{context6} displays the horizontal flows over a wide field of view measured using the
\emph{LCT-Doppler} method and then averaging  the resulting velocities over 3~hours, before and after the 
filament eruption. 
Before the eruption we can clearly see the north--south stream parallel to and about  10\degr{} east
of the filament. This stream disturbs differential rotation and brings plasma and magnetic structures to 
the south. Although differential rotation tends to spread the magnetic lines to the east, the observed 
north--south stream tends to shear the magnetic lines. After the eruption, only a northern segment of the filament 
is visible and the north--south stream has disappeared.

 To quantify the evolution of the horizontal flow before and after the filament eruption, 
we computed the change in the direction of the velocity vectors.
 The noise discussed in sect. 4.1, which can affect the direction of small magnitude vectors, 
tends to reduce the correlation between flows. In order to mitigate this error, we computed the magnitude 
weighted cosine of the direction difference (as in \v{S}vanda et al., 2007), which is robust to 
the presence of noise. This quantity is given by the formula:

\begin{equation}
\rho_{\rm W} = \frac{\sum |{\mathbf a}| \frac{|{\mathbf a} \cdot {\mathbf b}|}
{|{\mathbf a}||{\mathbf b}|}}{\sum |{\mathbf a}|},
\end{equation}
where $\mathbf a$ and $\mathbf b$ are vector fields, ${\mathbf a} \cdot 
{\mathbf b}$ is a scalar multiplication and $|{\mathbf a}|$ is the magnitude. 
The closer this quantity is to 1, the better the alignment between two 
vector fields.

Figure~\ref{context61} displays the magnitude-weighted cosine map computed between flows before and after the
eruption of the filament shown in Fig.~\ref{context6}. This map was computed by using a sliding 
window with a size of 8.8\arcsec on a side (41.2\arcsec being the side of the data plot).  
 The  magnitude weighted cosine map (Fig.~\ref{context61}) reveals that changes 
in the vicinity of the north--south velocity stream  are significant, while the horizontal flow field in the remainder 
of the field of view is more stable. Although variations in the flow field were expected to be more or less random 
over the field of view, we observe in that particular case that most of the variations between before and after of 
the filament eruption are located in the north--south stream. The disappearance of the  north--south stream after 
the eruption could be linked to the eruption or to a natural evolution of the photospheric flows.

\begin{figure*}
\resizebox{9cm}{!}{\includegraphics{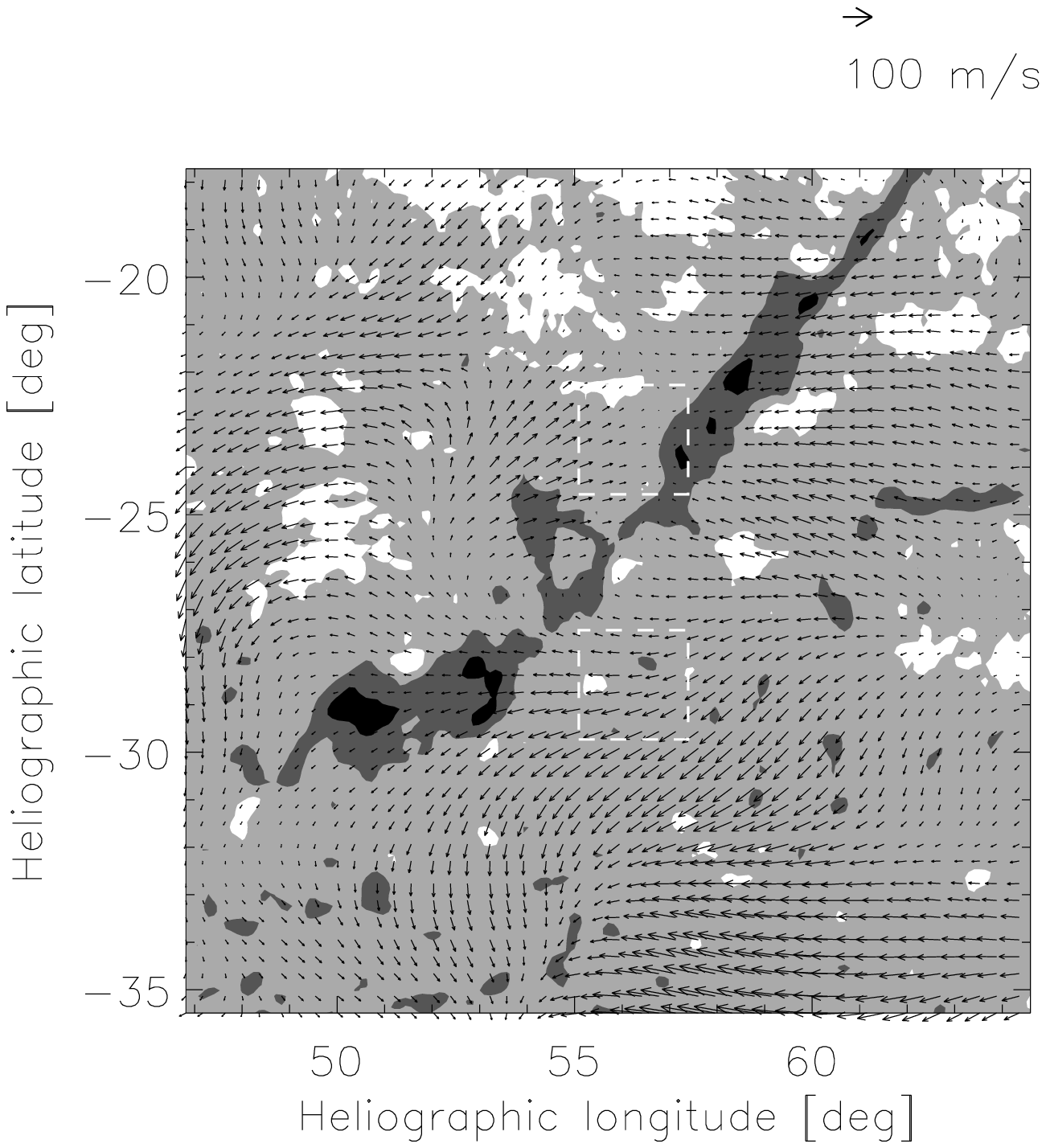}}
\resizebox{9cm}{!}{\includegraphics{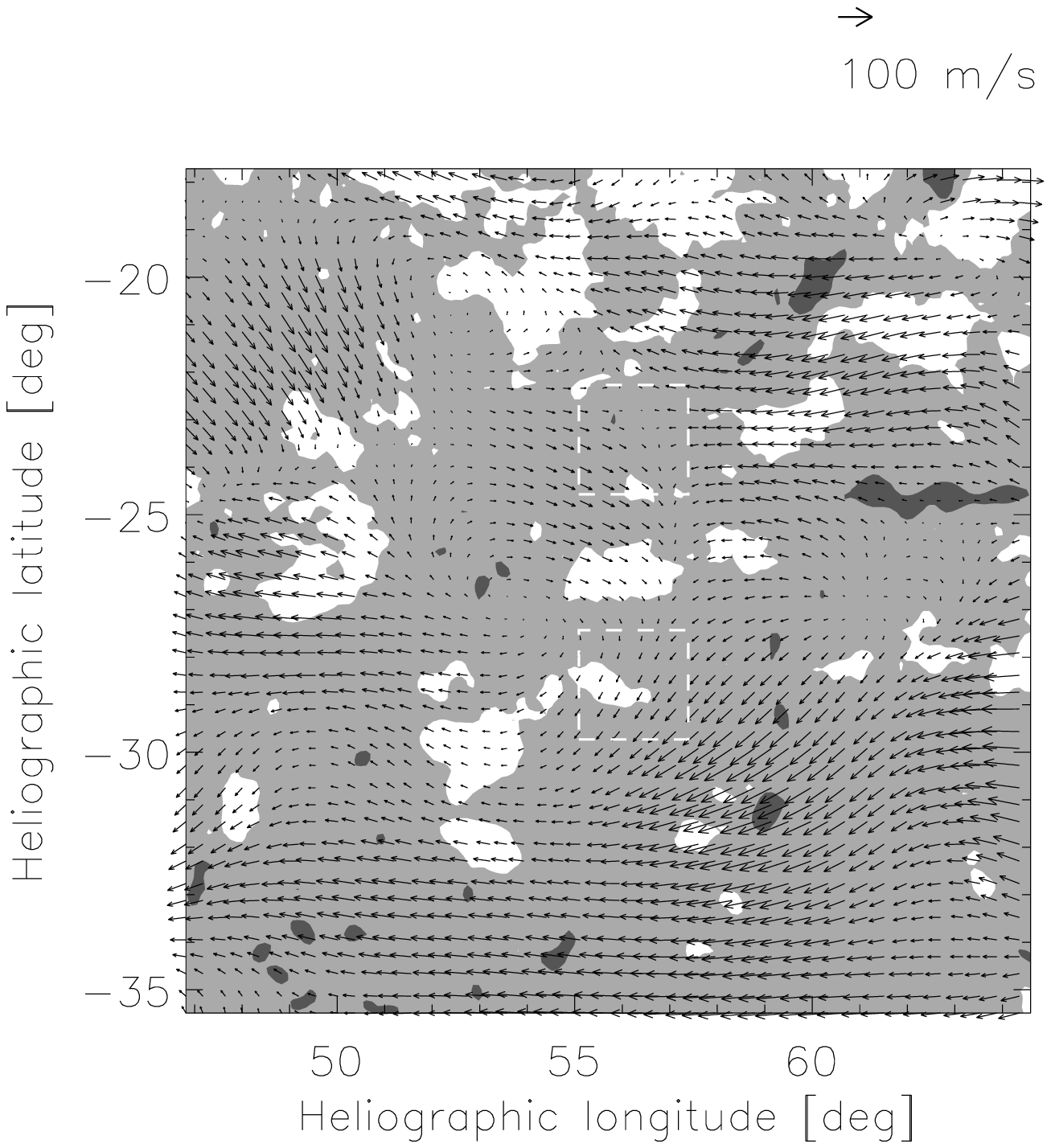}}\\
\caption[]{ A 3-hour average of the flow field in the immediate vicinity of the starting point 
($l=56^\circ$, $b=-26^\circ$) before (left) and after (right) the filament eruption. The filament observed 
by ISOON on 2004 October 7 at 15.20~UT is superimposed. The dashed boxes denote areas used for the zonal shear 
calculation.}
\label{context9}
\end{figure*}

 Figure~\ref{context9} shows the flow field in more detail at the site where the eruption starts.
 The shear in the zonal component at the point where the eruption starts ($l=56^\circ$, $b=-26^\circ$ 
in Carrington coordinates) is clearly visible and exits before and after the eruption, although the 
shape of the apparent vorticity has changed. This location corresponds to the area of upflow observed 
in the Meudon H$\alpha$ Dopplergram (Fig. 8 in Paper I).

\begin{figure}
\centerline{\psfig{figure=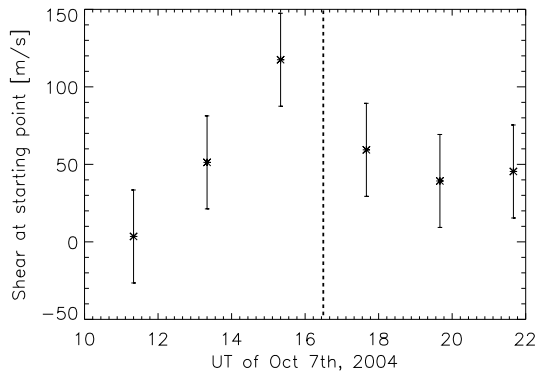,width=7cm}}
\caption[]{ The evolution of the velocity shear in zonal components in time.
 The eruption of filament took place at 16:30 UT}
\label{context11}
\end{figure}

We defined the shear as a difference between the mean zonal component $v_x$  in the area just  North 
and just South of the point at which the eruption appeared to start. We obtained the mean zonal flow by
averaging over boxes 2.3\degr{} on a side located 2.9\degr{} North and South of this point.
 The evolution in the shear velocity computed as the difference between the mean flow in the two boxes as a function of
time is shown in Fig.~\ref{context11}. Six 2-hour averages of the flow fields were used to create this figure. The 
error bars come from estimating the noise measured on the basis of synthetic data from \v{S}vanda et al.~(2006). 
One can see that the shear velocity increased before the eruption
and decreased after the eruption. One hour before the eruption, the shear reached the value of
(120$\pm$15)~m\,s$^{-1}$ over a distance of 5.2\degr{} (62\,000~km in the photosphere).
After the filament eruption, we observed the restoration of ordinary differential rotation below 30\degr{} south.

\section{Evolution over 6, 7, and 8 October, 2004}

\begin{figure*}
\resizebox{0.49\textwidth}{!}{\includegraphics{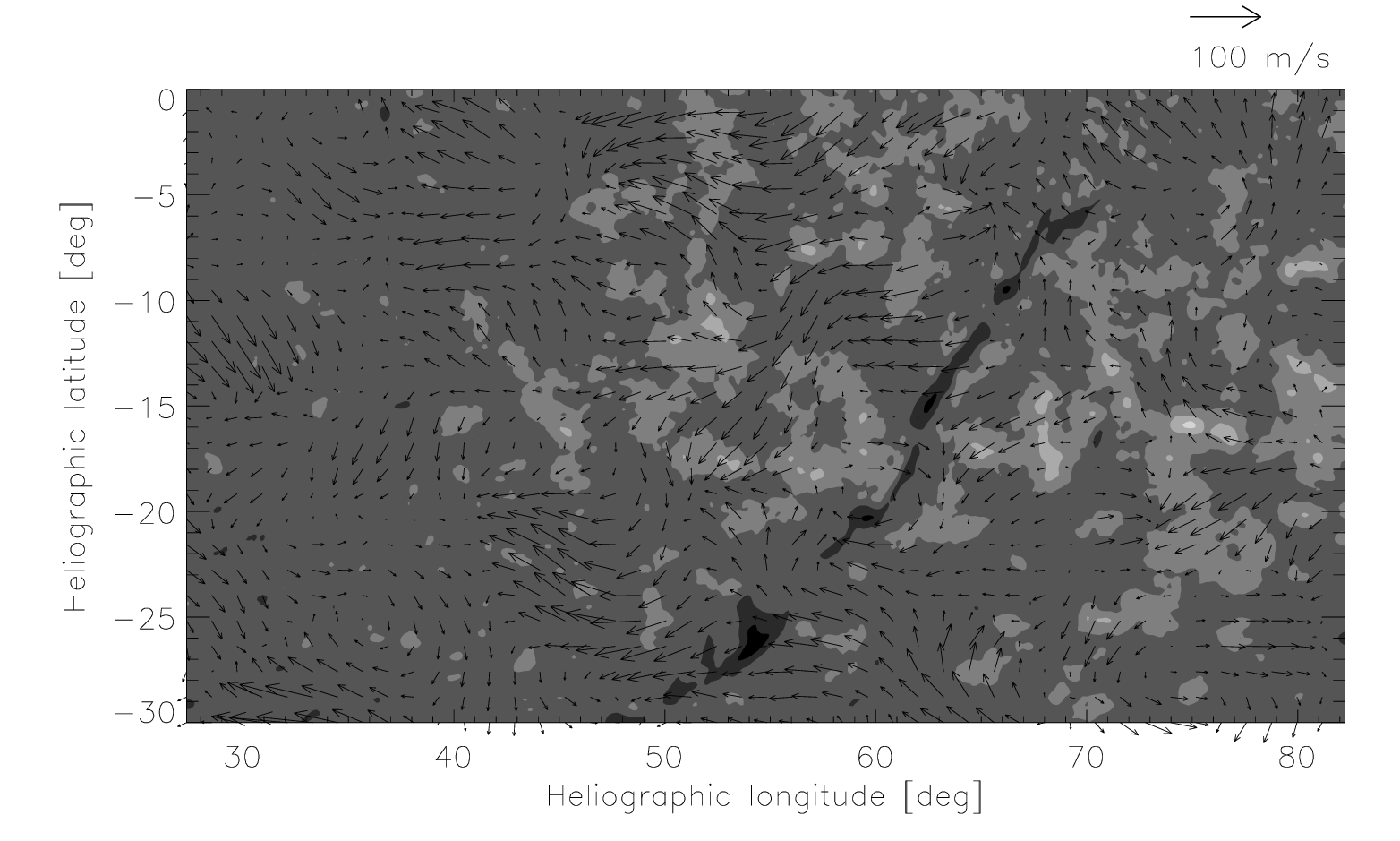}}
\resizebox{0.49\textwidth}{!}{\includegraphics{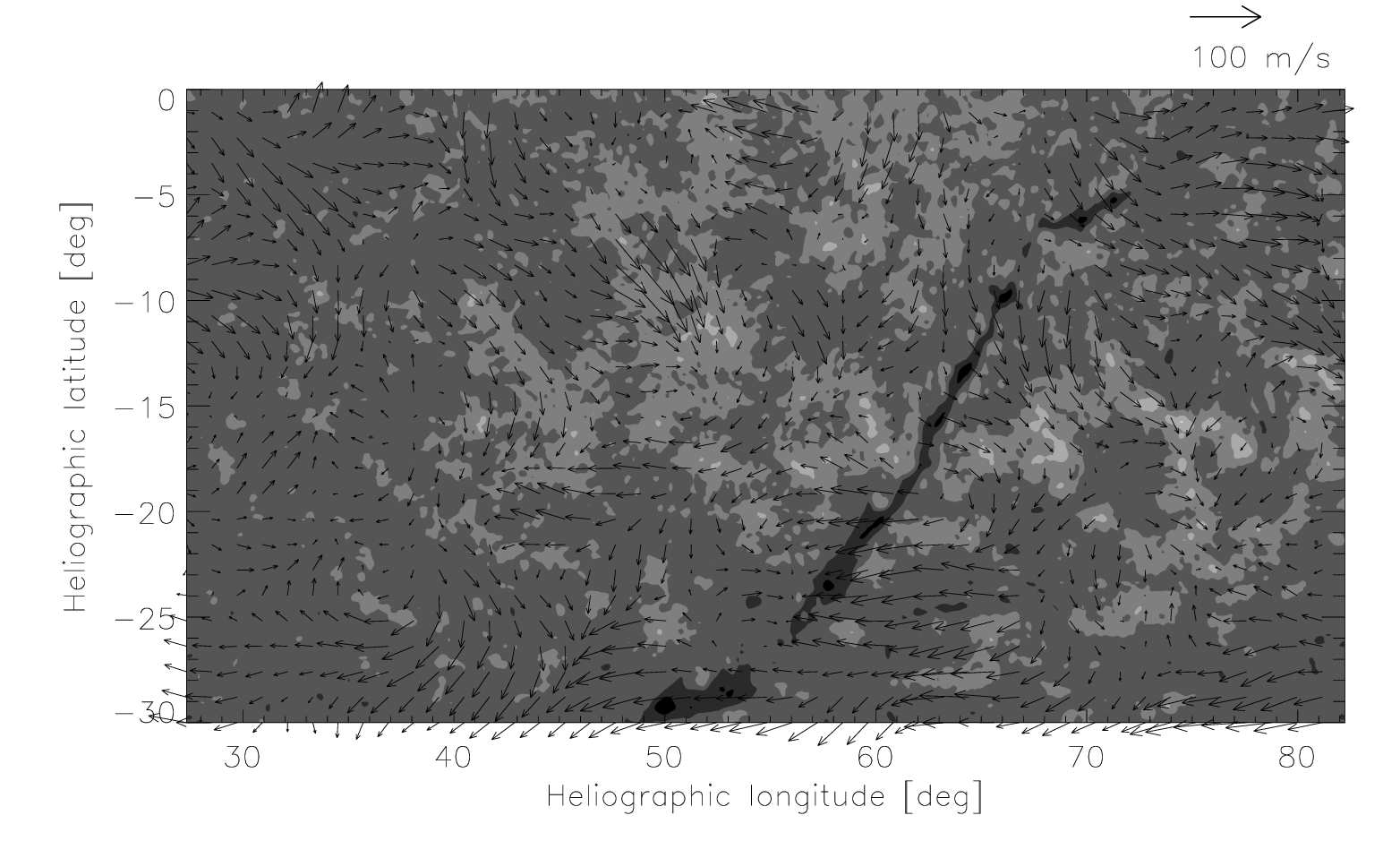}}\\
\resizebox{0.49\textwidth}{!}{\includegraphics{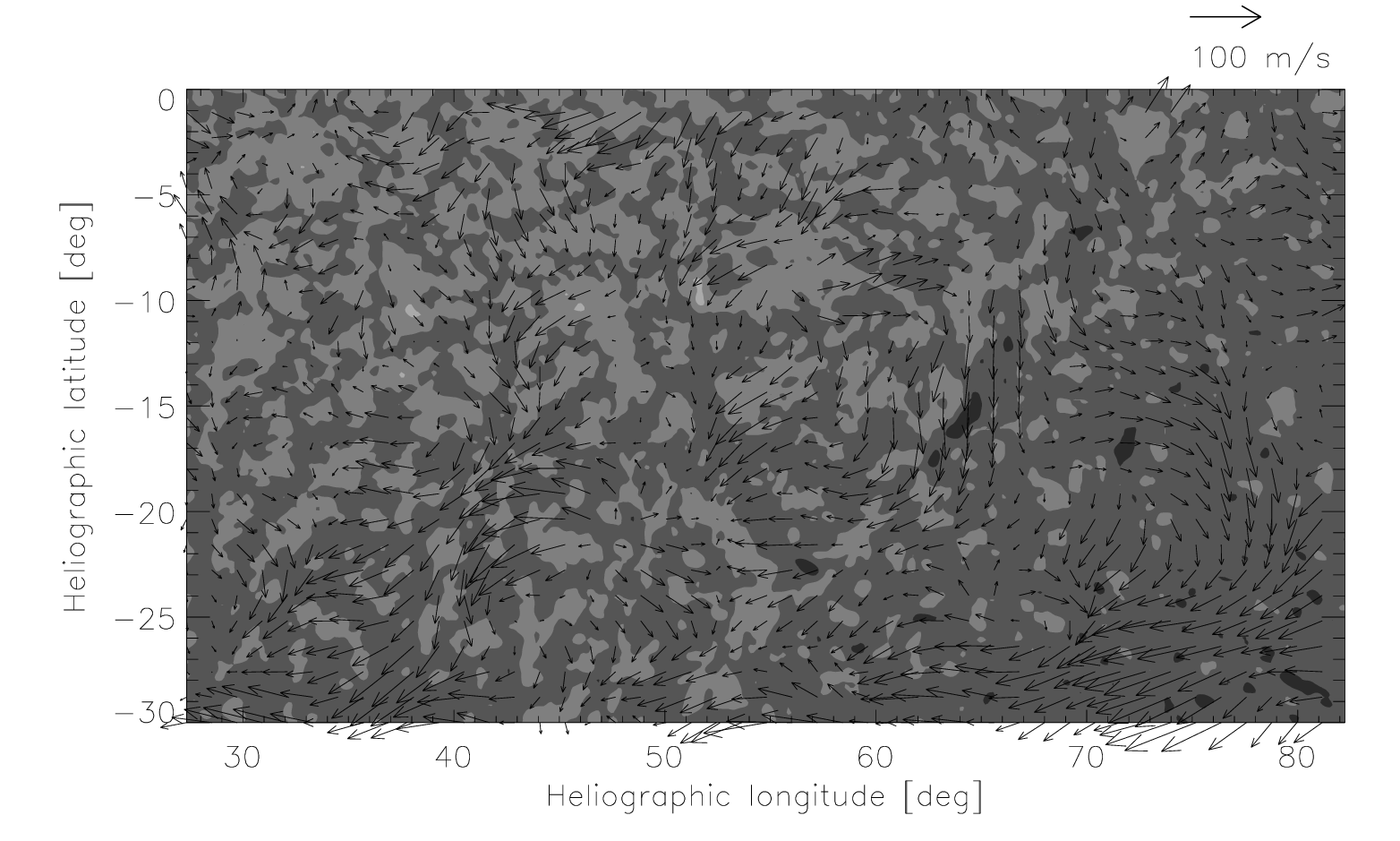}}
\resizebox{0.49\textwidth}{!}{\includegraphics{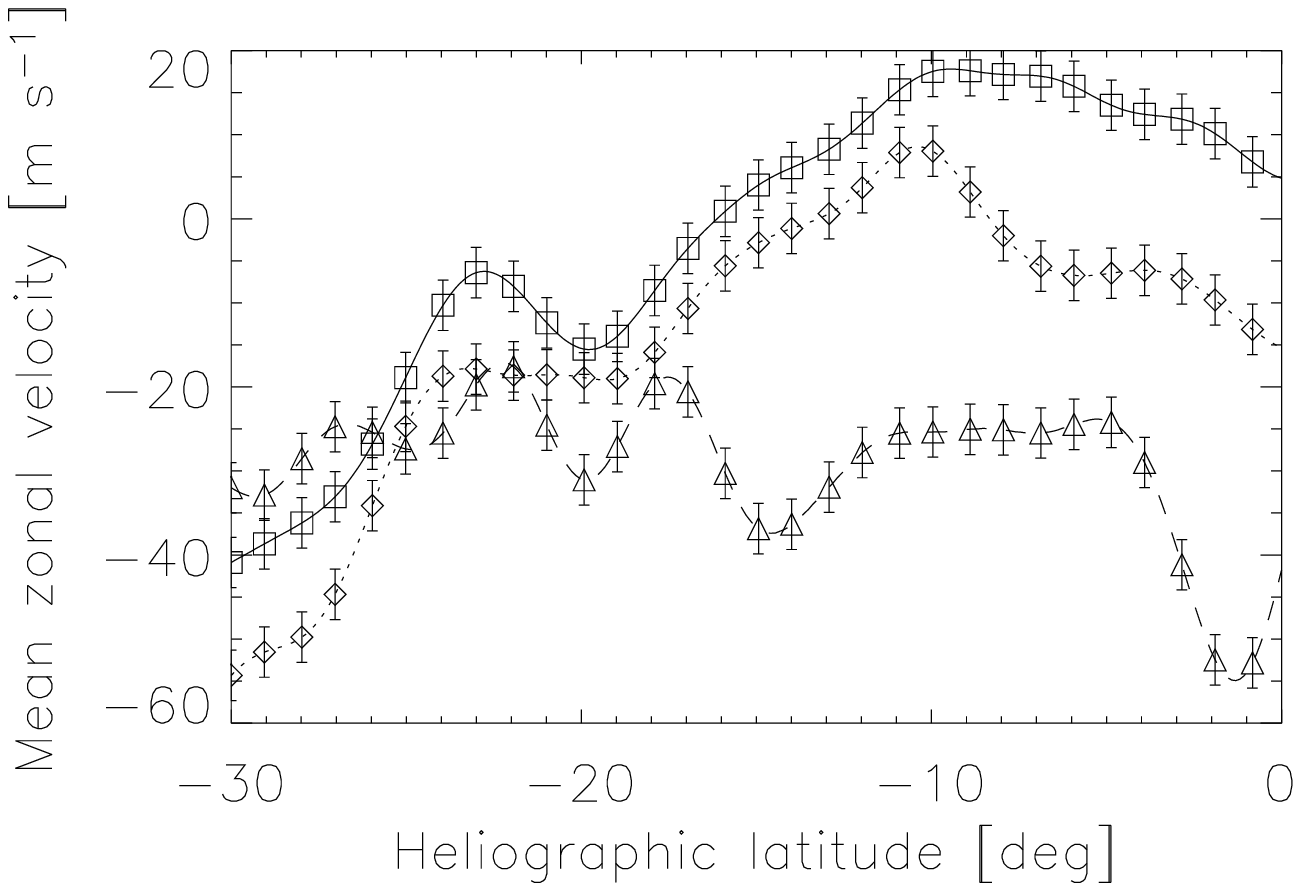}}
\caption[]{Horizontal flow evolution over three days 6 (\emph{upper left}), 7 (\emph{upper right}), 
and 8 (\emph{bottom left}) October, 2004. \emph{Bottom right} -- Mean zonal velocities, computed
between $-$35\degr{} and $+$20\degr{} in longitude and $-$30\degr{} to 0\degr{} in latitude.
\emph{Dashes (and triangles)} represent mean zonal velocity for the LCT applied  on Doppler 
on 2004 Oct 6. \emph{Solid lines (and squares)} represent mean zonal velocity for the LCT applied 
on Doppler on 2004 Oct 7. \emph{Dash dots (and diamonds)} represent mean zonal velocity for the 
LCT applied on Doppler on 2004 Oct 8. }
\label{context12}
\end{figure*}

In Fig.~\ref{context12} we compare the flow field in the filament region on the day of the eruption 
(October 7, 2004), with the flow fields on the preceding and following days. 
We see that the topology of the flows in this region changed over the three days:  the daily evolution of the mean 
zonal profiles is shown in Fig.~\ref{context12} (bottom right). The dashed line with triangles is the mean 
zonal velocity for 6 October, 2004. This profile is relatively flat probably due to the short time sequence as 
only 3 hours of data were available.
 The differential rotation profile for October 7 shows a secondary maximum around $-$23\degr{} in latitude, as 
discussed in Sect. 4.
 The October 8 profile exhibits similar trend, but with a smaller amplitude and an eastward
velocity for latitudes greater than 10\degr{}. The secondary maximum appears strongly reduced, indicating
restoration to a more regular differential rotation pattern in that zone.

 One day before the eruption, shear began to form at the site where the filament eruption is triggered
 ($l=56^\circ$, $b=-26^\circ$ in Carrington coordinates). The north--south stream is also visible. Both 
phenomena may store free energy in the coronal magnetic field configuration. The topology of the flow and 
the stream are different a day after the filament eruption suggesting that, after the disappearance of the southern 
part of the filament the conditions in the photosphere below, the filament became more relaxed. This may suggest 
the mutual coupling of the photospheric flow and the configuration of the coronal magnetic field. To confirm 
this idea, high-cadence high-resolution images and magnetograms covering the eruption time would be needed.

\section {Discussion and conclusion}

 Filaments or prominences are important complex structures of the solar atmosphere because
they are linked to  CMEs, which can influence the Earth's atmosphere and near-space environment. Surface motions acting
 on pre-existing coronal fields play a critical role in the formation of filaments. They appear to reconfigure 
existing coronal fields, by twisting and stretching them, thereby depositing energy in the 
topology of the coronal magnetic field. Photospheric motions can also initiate coronal magnetic field 
disruption. Surface motions play an important role in forming Type~B filament  that are located between young and 
old dipoles and have a long, stable structure. This class of filaments (Type~B) requires surface motions to gradually 
reconfigure preexisting coronal fields.

In a previous study (Paper I), we removed all the large-scale flows in order 
to focus on smaller scale flow, such as mesogranulation and supergranulation. In this paper we have retained the 
large-scale flows in order to study their influence on the triggering of a filament eruption.

The three  different methods used to estimate horizontal flows, while exhibiting small differences, provided a
consistent picture of the general trends of the flow patterns. 
 The \emph{LCT-B} method gave smaller velocity amplitudes but the flow directions agreed quite  well
with the other methods. The smaller amplitude occurs because of the small number of the small isolated magnetic 
elements and the smoothing effect of the spatial window used by this method. This leads to smoothing the results 
of the \emph{LCT-B} method by  approximately a factor of two. This agrees with the previous results of Schuck (2006), 
who shows that this method is not suitable for determining accurately estimating of the magnetic footpoint velocities. 
Correlation coefficients comparing velocities from the  \emph{manu-B} and \emph{LCT-Doppler} methods are positive 
and significant. General trends are very similar; however, we see many local discrepancies in the measured flow fields 
due to the noise. This implies that magnetic elements (detected by the \emph{LCT-B} and \emph{manu-B} methods) do not 
necessary follow the plasma flow (detected by the \emph{LCT-Doppler} method).

The filament eruption started at about 16:30~UT at the latitude around $-$25\degr{} where the measurements of the
horizontal flows based on Dopplergram tracking show a modification of the slope in the differential rotation
of the plasma. This behaviour is not observed in the curves obtained by tracking the 
longitudinal magnetic field; both methods show a continuous slope in the differential rotation in the same place. 
This seems to be a consequence of the presence of a north--south stream along the filament position, which is 
easily measured by tracing Doppler structures and is only slightly visible in maps obtained by 
tracking of magnetic elements.

The observed north--south stream has an amplitude of 30--40~m\,s$^{-1}$. In the sequence of H$\alpha$ image that record 
the filament's evolution, the part of the filament, which is in the north--south stream, 
is rotated in a direction compatible with the flow direction of the stream. This behaviour suggests that the foot-points 
of the filament are carried by the surface flows. The influence of the stream is strengthened by  
differential rotation. We should keep in mind that the filament extends from $-$5\degr{} to $-$30\degr{} 
in latitude and that the northern part of the filament is subjected to a larger rotation than the southern 
part. 

The  north--south stream, along with contribution from differential rotation causes the stretching of the coronal 
magnetic field in the filament and therefore contributes to destabilizing the filament. The topology of the north--south 
stream changed after the filament eruption, nearly vanishing.

We have measured an increase in the zonal shear at the site where the filaments eruption begins, before and its 
sudden decrease after the eruption. This result suggests that the shear in the zonal 
component of the flow field is the most important component of the surface flow affecting the stability 
of the coronal magnetic field, and it can lead to its eruption, which in turn can drive active phenomena 
such as ribbon flares and CMEs. This evolution of the shear in the flow field is probably related to the 
re-orientation by 70\degr~ (or 110\degr) of the transverse field after the eruption seen in the daily vector 
magnetograms obtained with THEMIS (Paper I)).

All of the features observed in the topology of the horizontal velocity fields at the starting-point site
could contribute to destabilizing the filament, resulting in its eruption.
  The present study has only examined the flows in the vincinity of a single filament. From our data, 
we propose  that the stability and evolution of the filament are influenced by surface flows that carry the footpoints 
of the filament. In addition, Dud\v{i}k et al. (2007) constructed a linear magnetohydrostatic model of the northern part
of the filament. His models show that the shape of the dipped field lines of the central part of the filament footpoints
closely resemble the shape of the underlying,  nearby polarities. This suggests a reconnection could be taking 
place between the flux of the incoming parasitic polarity and the ``native'' flux of the weak polarities dominant 
in that part of the filament channel.

Filaments, or prominences, are important complex structures of the solar atmosphere. Several mechanisms 
 are probably involved in the filament eruptions: the action of surface motions
 to create or increase the helicity of the flux rope (van Driel-Gesztelyi 2005, Romano 2005), reconnecting 
field lines in the corona (Mackay and van Ballegooijen, 2006a), the chilarity evolution of the barbs 
(Su et al. 2005), and oscillations  of the filament (Pouget 2006), etc. 
  The coronal magnetic field is generally thought to be anchored in the photosphere, and flux 
transport on the solar surface (Wang et al. 1989) is the natural mechanism to explain the evolution of 
filament. Recent models of the large-scale coronal structure (Mackay and van Ballegooijen, 2006a) consider the action
of the large-scale surface motions, such as differential rotation, meridional flow, and surface diffusion 
(supergranular).  Recent analysis of the near-surface flows computed from Doppler imaging, provided
by the MDI/SOHO instrument, reveals the shearing flow aligned with the neutral line (Hindman et al. 2006).
Our present observation indicates that large-scale surface flows  are structured 
(not uniform), showing areas of divergence or stream flows that should be taken into account in the 
numerical simulations. 
  
A better understanding of the mechanisms that lead to filament eruptions requires simultaneous 
multi-wavelength and multi-spatial resolution observations (both high resolution of the filament and low resolution 
of the full sun) over a wide range of latitudes. 
Indeed, our previous works showed that different phenomena are observed at high resolution, such as magnetic 
reconnection close to the starting location of the filament eruption (Paper I). In this paper, observing a
larger area at lower resolution, we showed that, at the same location where the filament first begins to erupt, 
there is a steep gradient in differential rotation, a north-south stream, and a shear in the zonal component.

 The next step in our study of the filament eruptions will be to examine the evolution of the extrapolated
coronal field from photospheric longitudinal magnetograms to determine whether the effects of surface motions 
on coronal fields play a critical role in causing filament eruptions.

\begin{acknowledgements}

This work was supported by the Centre National de la Recherche
Scientifique (C.N.R.S., UMR 5572, and UMR 8109), by the Programme National Soleil Terre (P.N.S.T.), by
the Czech Science Foundation under grant 205/04/2129, and by ESA-PECS under grant No.~8030. SOHO is a mission 
of international cooperation between the ESA and NASA.
This work was supported by the European commission through the RTN programme (HPRN-CT-2002-00313).
We wish to thank ISOON/O-SPAN, SOHO/MDI, SOHO/EIT teams for their technical help.
\end{acknowledgements}

\end{document}